\documentclass[sigconf]{acmart}

\AtBeginDocument{%
  \providecommand\BibTeX{{%
    \normalfont B\kern-0.5em{\scshape i\kern-0.25em b}\kern-0.8em\TeX}}}

\acmConference[IDE '24]{Make sure to enter the correct
  conference title from your rights confirmation email}{April 20,
  2024}{Lisbon, Portugal}
\acmBooktitle{IDE'24: ICSE Workshop on Integrated Development Environments,
 April 20, 2024, Lisbon, Portugal} 
\acmISBN{978-1-4503-XXXX-X/18/06}

\copyrightyear{2024}
\acmYear{2024}
\setcopyright{acmlicensed}\acmConference[IDE '24]{2024 First IDE Workshop}{April 20, 2024}{Lisbon, Portugal}
\acmBooktitle{2024 First IDE Workshop (IDE '24), April 20, 2024, Lisbon, Portugal}
\acmDOI{10.1145/3643796.3648445}
\acmISBN{979-8-4007-0580-9/24/04}
\begin{document}

\title{Lessons from a Pioneering Software Engineering Environment: Design Principles of Software through Pictures}

\author{Anthony I. (Tony) Wasserman}
\email{tonyw@acm.org}
\orcid{0000-0003-3841-8085}
\affiliation{%
  \institution{Software Methods and Tools}
  \city{San Francisco}
  \state{CA}
  \country{USA}
  \postcode{94131}
}

\renewcommand{\shortauthors}{Wasserman}

\begin{abstract}
  This paper describes the historical background that led to the development of the innovative Software through Pictures multi-user development environment, and the principles for its integration with other software products to create a software engineering environment covering multiple tasks in the software development lifecycle. 
\end{abstract}




\received{1 December 2023}
\received[accepted]{11 January 2024}

\maketitle

\section{Introduction and Background}
The first programming environments, supporting development and execution of a program written in a high-level language, were built in the late 1960s and early 1970s. These included Dartmouth BASIC 
\cite{Kemeny85},
WATFOR and WATFIV 
\cite{Cress68}, 
INTERLISP
\cite{Teitelman81}, with Smalltalk-80 following in 1980 
\cite{Goldberg83}. 
All of them allowed programmers, mostly students in the first two cases, to use only the programming environment for editing, coding, and execution of their programs.  Such tools came to be known as Integrated Development Environments (IDEs).
Interest in and use of programming environments grew rapidly following the release of the inexpensive Turbo Pascal in 1983 for DOS, originally developed by Anders Hejlsberg at Borland 
\cite{Turbo24}. 
Figure 1 shows a Turbo Pascal screen from Release 7.1 (1992).
\begin{figure}[h]
  \centering
  \includegraphics[width=\linewidth]{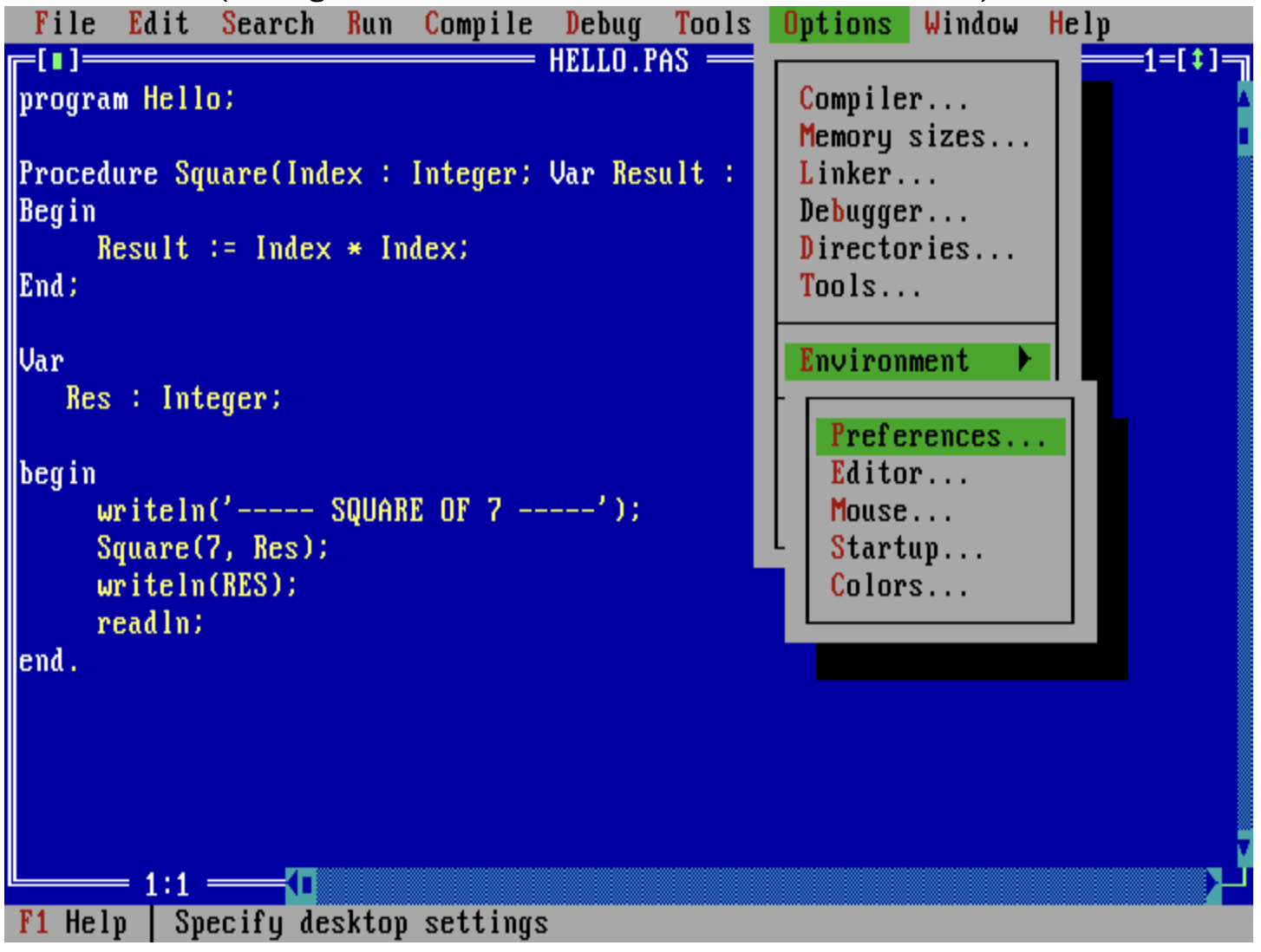}
  \caption{Turbo Pascal 7.1 screenshot}
  \Description{Turbo Pascal 7.1 screenshot}
\end{figure}

The field of software engineering emerged at the end of the 1960s, bringing attention not only to the programming task, but also to the broader topic of software development processes and tools covering aspects other than coding. When the US Department of Defense sponsored the design of the Ada programming language, they also sponsored creation of the “Stoneman” document, which specified requirements for Ada Programming Support Environments (APSE)
\cite{Stoneman80}. 
Stoneman described the APSE as a coordinated and complete set of tools which is applicable at all stages of the system life cycle, and that communicate mainly via the database, which stores all relevant information concerning a project throughout its life cycle. This notion of a shared repository made it possible to think about both individual and multi-user APSE’s.

\section{User Software Engineering}
At the time, the author was working on the User Software Engineering (USE) project 
\cite{Wasserman79}
\cite{Wasserman82}
\cite{Wasserman86}
Much of the early work on software engineering focused on programming methodology, emphasizing programming style, modular development, and top-down design. By contrast, the USE approach introduced external design, i.e., user interface design, into its methodology. 
The USE approach represented interactive information systems with a hierarchy of state transition diagrams, where nodes represented system output to the user, with  transitions (arcs) based on user inputs and possibly triggering actions. 
The USE approach was accompanied by a tool suite, including a rapid prototyping tool (RAPID/USE)
\cite{WassShew82}, 
a relational DBMS (Troll/USE) 
\cite{Kersten81}, and PLAIN 
\cite {VDRiet81}, a Pascal-like programming language with relations as a data type. This combination, known as the Unified Support Environment, was later joined by a graphical transition diagram editor to replace the original text-based language for describing transition diagrams.
Following the lead of BSD Unix 
\cite{Leffler81}, these tools were separately distributed through the University of California with a BSD license beginning in 1980.

\section{User Software Engineering}
In response to user requests for technical support of these tools, the author founded Interactive Development Environments, Inc. (IDE) in 1983. At the time, there were a growing number of graphically-based software design methods, including Structured Design 
\cite{Stevens74}, Entity-Relationship Diagrams 
\cite{Chen76}, HIPO, and many others. With many of the foundations already in place with the Transition Diagram Editor running on Sun workstations, IDE grew by supporting several of these other notations. 
The tools weren’t just for drawing, but also captured the semantics of the various notations. As a result, it was possible to check the consistency of models created with the editors, and to generate artifacts such as database schemas, data dictionaries, and  code skeletons from the models. That led to the name Software through Pictures (StP) 
\cite{Wasserman87}. Figure 2 shows a Structure Chart with a popup menu for further refinement.
\begin{figure}[h]
  \centering
  \includegraphics[width=\linewidth]{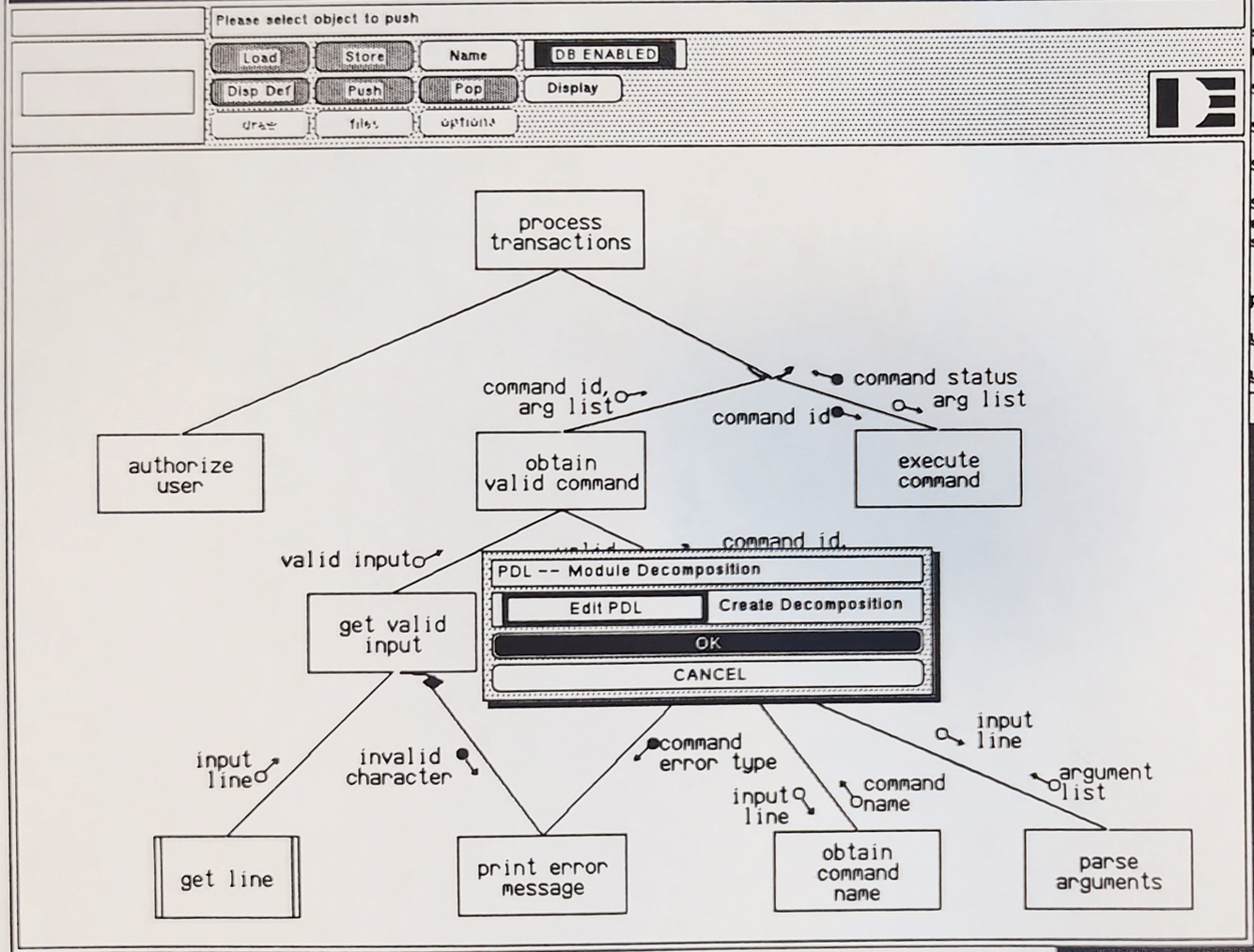}
  \caption{A Structure Chart diagram in StP}
  \Description{A structure chart diagram displayed in Software through Pictures.}
\end{figure}

\subsection{Innovative Aspects}
The design of Software through Pictures was strongly influenced by the design principles of Unix. StP was not a monolithic system, but rather a collection of modules (more than 50) that communicated with one another through a shared relational DBMS and procedure calls. All of the interfaces and file formats were documented and made available to users, including the schema of the database, making StP as open as possible, and available to others for tool integration.
The use of a relational DBMS, following the APSE concept, made StP into a multiuser environment. StP took advantage of the Network File System (new at the time) 
\cite{Sandberg85} and the X Windows System 
\cite{Scheifler86}. The result was that StP worked in a heterogeneous network environment with different brands of workstations.

\subsection{Adoption and Enhancement}
StP was commercially launched in November, 1984, and was regularly updated for more than 15 years, supporting new design notations and new platforms, including a Windows version for UML 
\cite{Booch05}. That product eventually became OpenAmeos in the early 2000s, with source code available.
Another important direction was the ability of StP to work with other software products. The first such product, available in 1992, was the C Development Environment, which connected StP with the Sabre-C (later renamed Centerline) IDE and the FrameMaker publishing system, allowing StP users to export their design artifacts for documentation and coding. (Unfortunately, there was no support for round-trip engineering.) Later versions supported both C++ and Ada with different IDEs.

\section{Tool Integration Principles}
This work led to development of a framework for describing tool integration.
\cite{Wasserman90}
Figure 3 shows four dimensions of tool integration in a software engineering environment with multiple tools presentation, platform, data, and control. 
\begin{figure}[h]
  \centering
  \includegraphics[width=\linewidth]{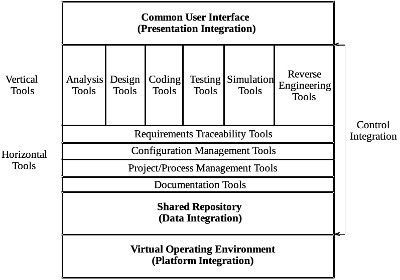}
  \caption{Dimensions of Tool Integration}
  \Description{Illustration of the different types of software tool integration in a software engineering environment}
\end{figure}
There is a  fifth dimension – process integration – that can describe the development process underlying the use of such tools. Ideally, tool integration occurs along multiple dimensions to create a well-integrated SEE.

ECMA adopted a similar model, shown in Figure 4, developed at HP
\cite{Earl90}, and known as the “toaster” model. 
\begin{figure}[h]
  \centering
  \includegraphics[width=\linewidth]{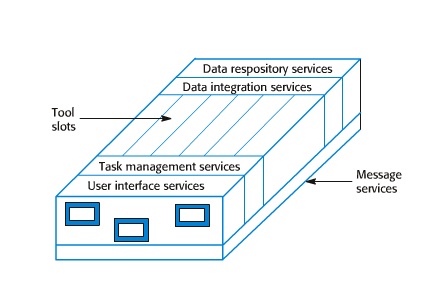}
  \caption{The ECMA Reference Model for SEEs}
  \Description{The ECMA Reference Model for software engineeting environments}
\end{figure}

\section{Conclusion}
The work described here, which goes back more than 50 years, has served as an important foundation for the creation of today’s sophisticated IDEs, which are now being further enhanced by the advent of artificial intelligence tools such as CoPilot
\cite{CoPilot24}. Today, software developers use many different tools and libraries to build their applications, making the need for powerful integration mechanisms more critical than ever before. The foundational approaches of Unix and Software through Pictures remain valuable for this work.

\bibliographystyle{ACM-Reference-Format}
\bibliography{IDE24}

\end{document}